\documentstyle[prl,aps,preprint,psfig]{revtex}

\begin{document}

\title{Josephson Effect in Unconventional Superconductors}

\author{Supriyo Datta and  Manoj P. Samanta }
\address{School of Electrical and Computer Engineering and the MRSEC
for Technology Enabling Heterostructure Materials}
 
\address{Purdue University, West Lafayette, IN 47907-1285, USA.}
 
\date{\today}
 
\vspace{-0.3in}
\maketitle
 
\begin{abstract}
In this letter we present a simple relation, that allows one to predict
the Josephson current between two unconventional superconductors with
arbitrary band-structure and  pairing symmetry.
We illustrate this relation with examples of s-wave and d-wave
junctions and show a detailed numerical comparison with the phenomenological
Sigrist and Rice relation~\cite{Sigrist} commonly used to interpret
tunneling experiments between d-wave superconductors. Our
relation automatically accounts for `mid-gap states' that occur
in d-wave superconductors~\cite{Hu,Rainer} and clearly shows how
these states lead to a temperature-sensitive contribution
to the Josephson current for certain orientations of the junction.
This relation should be useful in exploring
the effects of disordered junctions, multiple atomic
orbitals and complicated order parameters~\cite{Mazin,Harlingen}.
\end{abstract}
 
\pacs{}

\section{INTRODUCTION}
\label{intro}

Phase-sensitive measurements involving Josephson junctions has recently 
become an important technique for probing the pairing symmetry of 
high-$T_c$ superconductors~\cite{Harlingen,Kirtley,Delin}. Several 
experiments point to the existence of $\pi$ junctions between high-$T_c$ 
superconductors~\cite{Harlingen,Kirtley}, strongly suggesting d-wave 
symmetry, though other possibilities cannot be ruled out completely. 
In this paper we present a simple expression (valid for junctions with
arbitrary pairing symmetry), that shows how a particular feature in the
density of states contributes to the critical current in a Josephson 
junction. For example, it is now well-known~\cite{Hu} that, for 
certain orientations, d-wave superconductors exhibit a midgap peak
in the surface density of states. Using our formulation, one can easily 
see with a `back of the envelope' calculation that this midgap peak
will lead to a large contribution to the critical current that will
decay inversely with temperature.

We write the current I between the two 
superconductors 1 and 2 in terms of the phase difference $\phi$ as
$I=I_c(\phi) \sin\phi$, where $I_c$ is given by
(f(E): Fermi function) :
\begin{eqnarray}
I_c=\frac{2e}{\pi\hbar} |M|^2 \int_{-\infty}^{\infty} dE f(E) J(E) \nonumber \\
J(E)={\rm Im}[g_2^{eh}g_1^{he}+g_1^{eh}g_2^{he}]_{(E+i\eta)} \label{eq:current}
\end{eqnarray}
In this paper we will only discuss the tunneling limit
 in which $I_c$ is independent of $\phi$ and  the pair 
correlation functions $g^{eh,he}_{1,2}$ can be approximated by their 
values at a free surface~\cite{note_strong}.
These functions are obtained by
evaluating the retarded Green function for the Bogoliubov-deGennes 
equation. Complicated band-structures and pairing symmetries can be
accounted for by an appropriate choice of the one-particle Hamiltonian 
H and pair potential $\Delta(r,r')$.

\begin{eqnarray}
\left(\begin{array}{c}
(E+i\eta)I-H \;\;\;\;\; -\Delta \;\; \\
-\Delta^\dagger \;\;\;\;\;\;\; (E+i\eta)I+H^*
\end{array}\right)
\left(\begin{array}{c}
G^{ee} \;\;\;\; G^{eh} \\
G^{he} \;\;\;\; G^{hh}
\end{array}\right)
\nonumber	\\
=
\left(\begin{array}{c}
\delta(r-r') \;\;\;\; 0 \\
0            \;\;\;\;   \delta(r-r')
\end{array}\right)		\label{eq:BdG}
\end{eqnarray}
The surface Green function 'g' appearing in eq~\ref{eq:current} is 
obtained from the full Green function $G(r,r')$ by setting $x=x'=0$,
$x$ being the normal to the surface:
\begin{eqnarray}
g^{eh,he}(y,z;y',z')=G^{eh,he}(r,r')_{x=x'=0}	\nonumber
\end{eqnarray}
In eq~\ref{eq:current}, we have assumed a one-dimensional model for 
which $g^{eh,he}$ as well as the coupling element M are purely numbers. 
A generalized version of this equation is presented at the end of 
this paper that can be used when $g^{eh,he}$ and M are matrices.
For our discussions in this paper the simple version in eq~\ref{eq:current}
is adequate since we will only consider planar junctions with s-wave
or d-wave order parameters, for which each transverse wave vector
$(k_y,k_z)$ decouples into an independent one-dimensional
channel. The generalized version is needed to handle
disordered junctions, multiple atomic orbitals and complicated 
multilayer order parameters~\cite{Mazin,Harlingen}.

The advantage of this
formulation is that it allows us to predict the Josephson
current from a knowledge of the pair correlation functions for the
individual isolated superconductors. From this point of
view eq~\ref{eq:current} is similar to the well-known formula for
the conductance ($R^{-1}$)
\begin{eqnarray}
\frac{R^{-1}}{e^2/\pi \hbar}=4 \pi^2 |M|^2 \rho_1 \rho_2
                                        \label{eq:resistance}
\end{eqnarray}
that is routinely used to obtain the density of states ($\rho_{1,2}$)
from experimental data. Indeed, the pair correlation functions
$g^{eh,he}(E)$ exhibit features similar to those in the density
of states and  can often be guessed intuitively as discussed
in section~\ref{discuss}.

We will not go into the derivation of eq~\ref{eq:current} which follows
from the Green function formalism~\cite{Arnold,Samanta_sc}.
However it can be seen easily that for ordinary s-wave superconductors
it yields the standard results. To establish contact with the literature,
we rewrite eq~\ref{eq:current} as a sum over the Matsubara energies
$\omega_n=(2n+1) \pi k T$ along the imaginary axis:
\begin{eqnarray}
I_c=\frac{2e k_B T}{\hbar} |M|^2  \sum_n [g_2^{eh} g_1^{he}+g_2^{eh} g_1^{he}]
_{(E=i\omega_n)}                        \label{eq:matsubara}
\end{eqnarray}
Making use of the property
$g^{eh}(i\omega_n)=g^{he}(i\omega_n)^*\equiv -F(i\omega_n)$ 
(which is true provided
$\Delta(r,r')=\Delta(r',r)$), eq~\ref{eq:matsubara} reduces to
\begin{eqnarray}
I_c=\frac{4e k_B T}{\hbar} |M|^2  \sum_n F_2(i\omega_n)F_1^*(i\omega_n)
                                \label{eq:reduce}
\end{eqnarray}
This result is often applied to s-wave junctions using the bulk pair
correlation function for F and can be combined with eq~\ref{eq:resistance}
to obtain the Ambegaokar-Baratoff
relation~\cite{Ambegaokar,Xu}. In d-wave (and some other unconventional)
superconductors, the surface pair-correlation functions can show
features like `midgap peaks' that are absent in the bulk. The point to note
is that eq~\ref{eq:current} ( or \ref{eq:matsubara} )
can be used for arbitrary pairing symmetry, {\it provided
we use the correct pair-correlation functions at the surface
rather than the ones in the bulk}.
Eqs~\ref{eq:current} and \ref{eq:matsubara} are equivalent and it is largely
a matter of taste which one we use. In this paper, we will use the former and
present our results in terms of the pair correlation functions and the
current spectrum along the real energy axis.

\section{QUALITATIVE DISCUSSION}
\label{discuss}

To use eq~\ref{eq:current} we need the pair correlation functions 
$g^{eh}$. In general it is fairly straightforward 
to obtain a numerical solution for complicated band-structure
and pairing symmetries~\cite{Samanta_sc}. But even without a
detailed quantitative solution, one may be able to guess $g^{eh}(E)$
from a knowledge of the density of states and use it to gain insight into the 
resulting effect on the Josephson current.
For example, for d-wave superconductors with small values of the 
misorientation angle $\alpha$ and transverse wave-vector ($k_y,k_z$),
$g^{eh}$ has the usual form encountered with s-wave 
superconductors~\cite{note:e_to_the_power}:
\begin{eqnarray}
g_{1,2}^{he}=g_{1,2}^{eh}\sim\frac{\Delta_{1,2}}{\sqrt{\Delta_{1,2}^2-E^2}}
					\nonumber
\end{eqnarray}
The current spectrum $J(E)$ is s-wave like and  the critical current has 
the usual temperature dependence.
But when the quantity $\sin 2\alpha \sin k_y a$ is large, $g^{eh}$
is dominated by a singularity at E=0 corresponding to the midgap 
peak in the density of states.
\begin{eqnarray}
g_{1,2}^{he}=-g_{1,2}^{eh}\sim \frac {iA_{1,2}}{E}	\label{eq:approx}
\end{eqnarray}
This results in a large contribution to the
critical current which, however, decays rapidly with temperature. 
To see this, we use eq~\ref{eq:approx} in eq~\ref{eq:current} to obtain
\begin{eqnarray}
J(E)=2 A_1 A_2 \frac{2 E\eta}{(E^2+\eta^2)^2}\;\; \rightarrow  
2 \pi A_1 A_2 \delta'(E) \;\; {\rm as } \;\; \eta\rightarrow 0 \nonumber
\end{eqnarray}
where $\delta'(E)$ represents the derivative of the delta function.
Hence from eq~\ref{eq:current}
\begin{eqnarray}
I_c=\frac{4e}{\hbar} A_1 A_2 |M|^2 {(-\frac{\partial f}{\partial E})}_{E=0}
=\frac{e}{\hbar} A_1 A_2 \frac{|M|^2}{k_B T} \nonumber
\end{eqnarray}
indicating that the contribution of the midgap peak to the critical 
current decays with temperature as ~$1/T$. The apparent divergence near 
T=0 can be eliminated either by including a finite $\eta$ to
represent phase-breaking effects or by going beyond the weak-coupling
approximation ($|M g^{eh}|<<1$)~\cite{note_strong}.
The recent results of ref~\onlinecite{Tanaka} for d-wave junctions
(which includes strong coupling effects) also indicate a similar anomalous 
temperature dependence of the midgap contribution to the critical current.

\section{numerical results}

For quantitative calculations, we evaluate the pair-correlation 
functions from the BdG equation using a two-dimensional tight-binding
lattice with a dispersion relation of the 
form($\mu/\Delta_0=15$, $t/\Delta_0=10$):
\begin{eqnarray}
E=-\mu+2t(1-\cos k_xa)+2t(1-cos k_ya)           \label{eq:tb}
\end{eqnarray}
and a pair  potential of the form:
\begin{eqnarray}
\Delta=\Delta_0 \cos 2\alpha (\cos k_xa-  \cos k_ya)  \nonumber	\\
+\Delta_0 \sin(2\alpha) (\sin k_xa \sin k_y a)    \label{eq:delta}
\end{eqnarray}
We set $\eta=.025$ which corresponds to a phase-relaxation time
of $\tau_\phi=2$ ps, if $\Delta_0$ is 20 meV. Fig 2 shows the 
bulk and surface correlation function for $\alpha=0$ and 30 degrees with
$k_ya=0,1$. If either $\alpha$ or $k_ya$ is zero, the midgap term (see 
eq~\ref{eq:approx}) is absent and the surface $g^{eh}$ looks
similar to the bulk $g^{eh}$. But
with both  $\alpha$ and $k_ya$ non-zero, the surface and bulk 
functions are completely dissimilar. The bulk pair correlation function 
exhibits two gaps corresponding to $+k_x$ and $-k_x$ respectively,
since they experience different  values of $\Delta$. The surface pair 
correlation function involves a non-trivial mixture of the two and
has a midgap peak as evident from the second of fig~\ref{fig:corr}b.

Fig 3a shows the current spectrum for a Josephson spectrum with
$\alpha_1=\alpha_2=30$ degrees. With $k_ya=0$, the spectrum is s-wave-like
with peaks close to $E=\pm\Delta_{1,2}$. But with $k_ya=1$ we have a
large midgap contribution. 
On integrating over $k_y$~\cite{note_integrate}, 
this midgap contribution completely dominates the usual
peak away from midgap, (see fig~\ref{fig:c}b). The midgap peak
has the form of a doublet function ($\delta'(E)$) and its contribution
to $I_c$  will disappear with increasing temperature as
discussed earlier.

Fig~\ref{fig:sigrist}a and b  show the critical current as a function
of the misalignment angles $\alpha_1$ and $\alpha_2$ with $kT=0$ and 
with $kT=.5 \Delta$ respectively. For 
comparison we have also shown the results from the phenomenological relation
given by Sigrist and Rice~\cite{Sigrist} (Fig ~\ref{fig:sigrist}c):
\begin{eqnarray}
I_c= I_{c0} \cos 2 \alpha_1 \cos 2 \alpha_2     \label{eq:Sigrist}
\end{eqnarray}
We note that in fig~\ref{fig:sigrist}a, the current 
has peaks at $\alpha_1=\pm \pi/4, 
\alpha_2=\pm \pi/4$ due to the large midgap peaks in $g^{eh}$, in sharp
disagreement with eq~\ref{eq:Sigrist}~\cite{note_midgap}. At higher
temperatures (fig ~\ref{fig:sigrist}b) the contribution of these peaks 
is reduced as explained earlier,
and the variation in the critical current looks closer to what we 
expect from eq~\ref{eq:Sigrist}. A full calculation of 
$I_c(T)$ requires a 
self-consistent evaluation of the pair potential~\cite{Rainer},
which we do not address in this paper. Using the self-consistent 
pair potential from ref~\onlinecite{Rainer}, we find that the current
contribution due to the midgap peak is suppressed somewhat, but its
effect should still be observable.

\section{generalization}

Finally we note that eq~\ref{eq:current} can be generalized to include
cases where the pair correlation functions as well as the matrix 
element M are matrices instead of pure numbers. The current spectrum is
then given by 

\begin{eqnarray}
J(E)={\rm Im} [{\rm Tr}[ g_2^{eh} M^+ g_1^{he} M 
+ g_2^{he} M^+ g_1^{eh} M]]_{(E+i\eta)} \label{eq:J}
\end{eqnarray}
instead of eq~\ref{eq:current}. We believe eq~\ref{eq:J} should be 
useful in calculating the Josephson current taking into account 
disordered junctions, multiple atomic orbitals, or complicated 
pair potentials. The required surface Green functions can be 
calculated numerically using an iterative scheme once the 
appropriate tight-binding Hamiltonian has been identified~\cite{Samanta_sc}.

\section*{ACKNOWLEDGMENTS}

It is a pleasure to thank Phil Bagwell for valuable discussions and 
encouragement. This work was supported by the MRSEC program of the 
National Science Foundation under Award No. DMR-9400415.

\begin{figure}
\centerline{\psfig{file=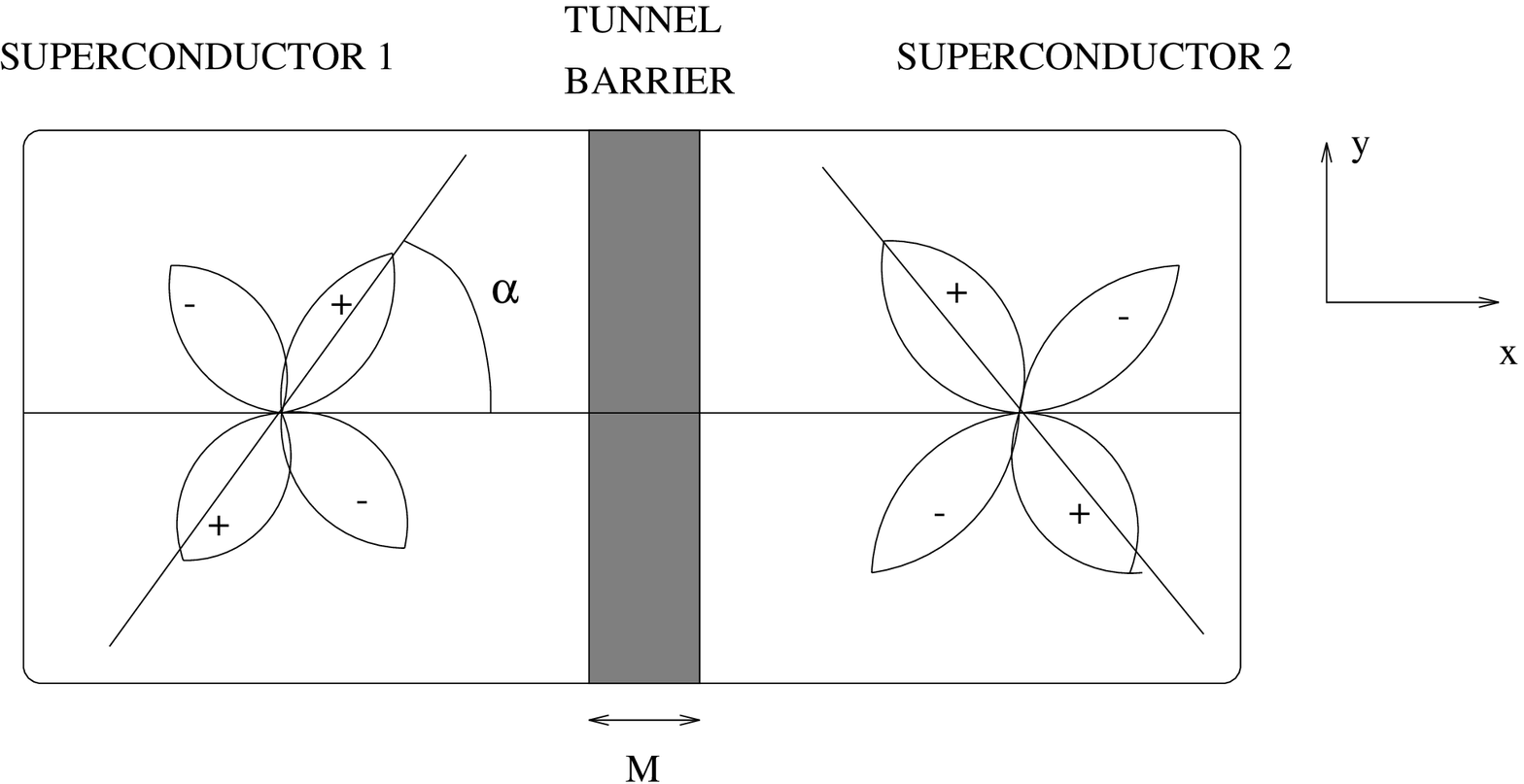,height=2 in}}
\vspace{.2in}
\caption{ Structure considered in this paper. M is the matrix element
linking superconductors 1 and 2. In case of d-wave, the pair potential
in each superconductor varies with the direction $\theta$
according to $\Delta(\theta)=\Delta_0 \cos(2\theta - 2 \alpha)$, where
$\alpha$ is the angle that the main lobe makes with the x-axis. }
\label{fig:structure}
\end{figure}
 
\begin{figure}
\centerline{\psfig{file=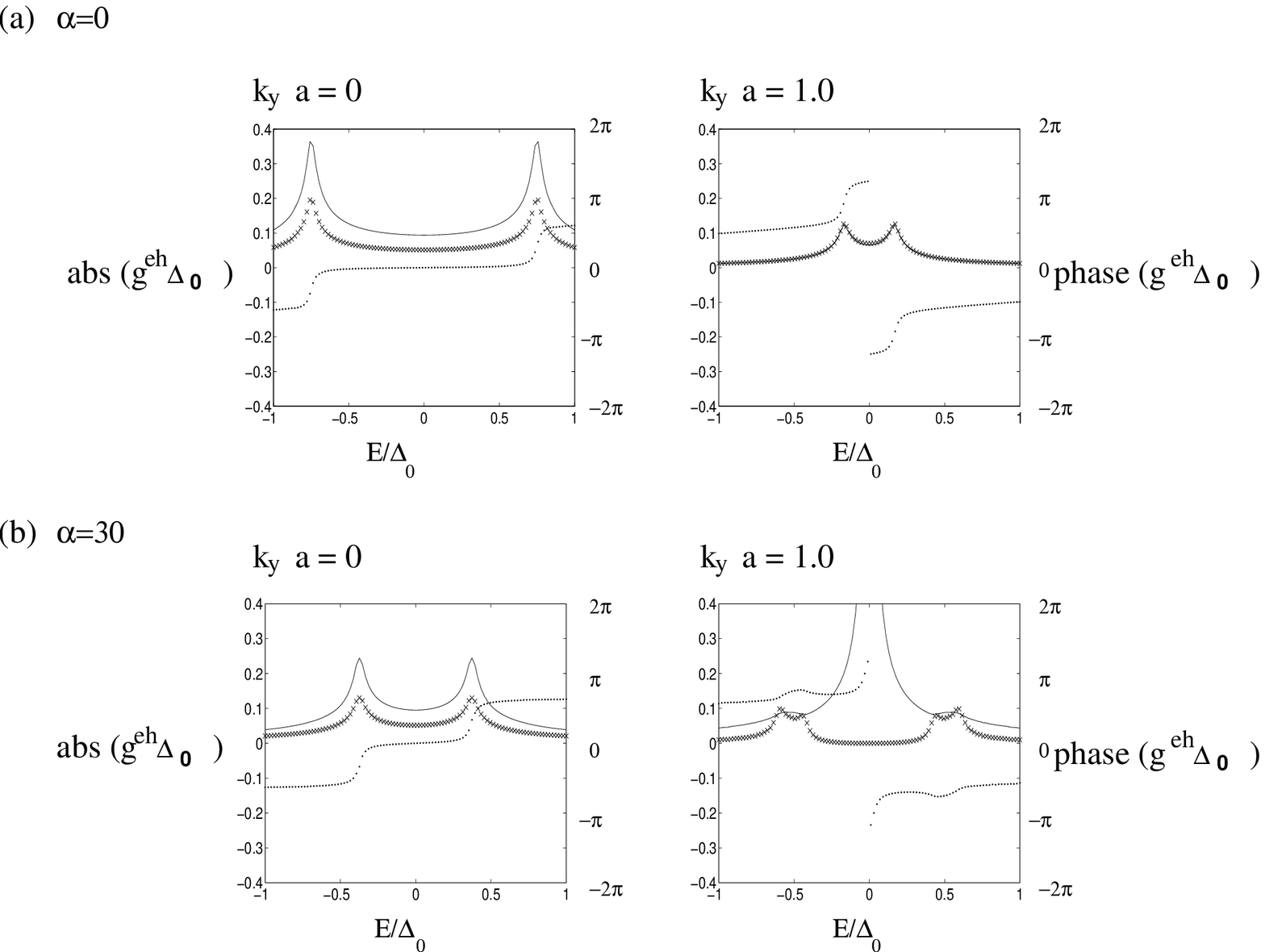,height=6 in}}
\vspace{.2in}
\caption{Magnitude and phase of the pair correlation
function $g^{eh}$ at a surface for $\alpha = 0$ degrees and
30 degrees with (i) $k_ya=0$ and (ii) $k_ya=1$. These correspond to
(i) $\theta=0$ and (ii) $\theta=50$ degrees respectively, since
$sin\theta=k_y/k_f$
and $k_fa=1.3$ . Also shown is the magnitude of the bulk pair
correlation function.  (----:surface(magnitude),
 .....: surface(phase), xxx: bulk(magnitude) ) }
\label{fig:corr}
\end{figure}
 
\begin{figure}
\centerline{\psfig{file=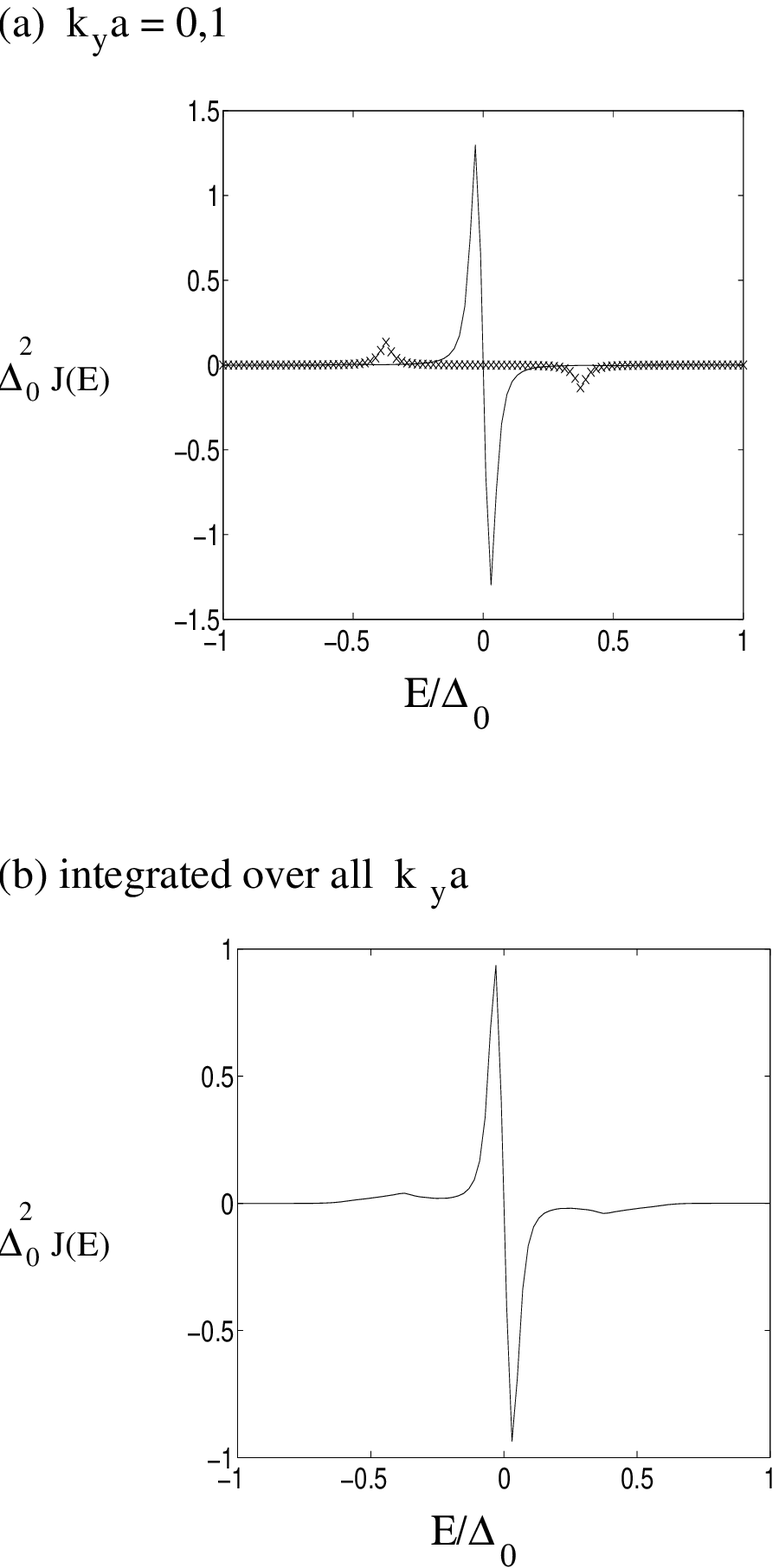,height=6 in}}
\vspace{.2in}
\caption{Josephson current spectrum for a d-wave junction having
$\alpha_1=\alpha_2=30$ degrees (a) with $k_ya=0,1$ (xxx : $k_ya=0$,
solid : $k_ya=1$) ,
(b) integrated over all values of $k_ya$.  }
\label{fig:c}
\end{figure}
 
\begin{figure}
\centerline{\psfig{file=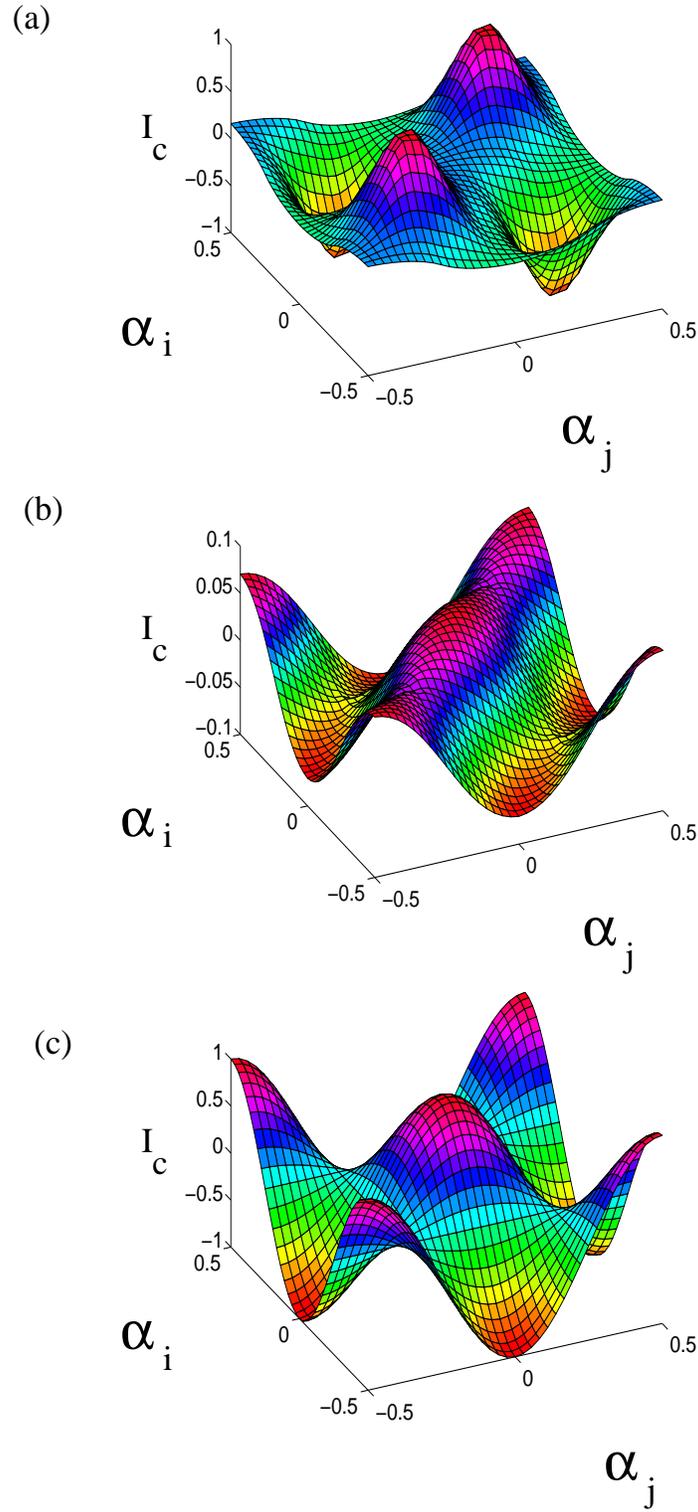,height=8 in}}
\vspace{.2in}
\caption{ Critical current (normalized) of Josephson junctions as a function
of the misalignment angles $\alpha_1$ and $\alpha_2$ (in units of $\pi$):
(a) $kT=0$,
(b) $kT=.5\Delta_0$ and (c) Phenomenological result based
on $I_c= I_{c0} cos(2 \alpha_1) cos(2 \alpha_2)$ }
\label{fig:sigrist}
\end{figure}


\begin{thebibliography}{10}
\bibitem{note_strong}
Eq~\ref{eq:current} is valid for strong coupling too, if we use the correct
  pair correlation functions taking the coupling into
  account~\cite{Samanta_sc}.
\bibitem{note:e_to_the_power}
Note that in evaluating $g^{eh}$ or $g^{he}$ we do not need to include in
  eq~\ref{eq:BdG} any overall phase factor $e^{i\phi}$ associated with the pair
  potential $\Delta$. Also, in general,
  $g^{eh}(E+i\eta)=[g^{he}(E-i\eta)]^\dagger$.
\bibitem{note_midgap}
However, in this paper (and in ref~\onlinecite{Tanaka}) only short clean
  junctions are considered. The mid-gap contribution may be suppressed for long
  or disordered junctions.
\end{thebibliography}
\end{document}